# Making Sense of the *Many Worlds Interpretation*

Stephen Boughn[1]
Department of Physics, Princeton University and
Departments of Astronomy and Physics, Haverford College

## Contents

## Preface

It has been 61 years since Hugh Everett III's PhD dissertation, *On the Foundations of Quantum Mechanics,* was submitted to the Princeton University Physics Department. After more than a decade of relative obscurity it was resurrected by Bryce DeWitt as *The Many Worlds Interpretation of Quantum Mechanics* and since then has become an active topic of discussion, reinterpretation, and modification, especially among philosophers of science, quantum cosmologists, and advocates of quantum decoherence and quantum computing. Many of these analyses are quite sophisticated and considered to be important contributions to physics and philosophy. I am primarily an experimental physicist and my pragmatic ruminations on the subject might be viewed with some suspicion. Indeed, Bohr's pragmatic *Copenhagen Interpretation* is often disparaged by this same cohort. Still, I think that my experimentalist's vantage point has something to offer and I here offer it to you. I gratefully acknowledge the inspiration I have drawn from Freeman Dyson and Marcel Reginatto; although, of course, they deserve no blame whatsoever for any foolish proclamations I might have made.

---

[1] sboughn@haverford.edu

1.  IN THE BEGINNING ….

I was beginning my second year of graduate school when Bryce DeWitt's article (DeWitt 1970), "Quantum mechanics and reality", appeared in *Physics Today*.  I had already begun to fret about the interpretation of quantum mechanics and so I read DeWitt's piece with great interest.  Two things about quantum mechanics had bothered me most.  I worried a great deal about the fundamental statistical aspect of the theory.  The deterministic Schrödinger equation governed the evolution of the wave function (probability amplitude) but as far as I could discern, quantum theory proper was silent on the outcome of any particular measurement; according to quantum mechanics, nothing ever "happens".  This worry was supposed to be allayed by Bohr's *Copenhagen Interpretation* that instructs us to interpret the quantum state in terms of the probability with which a *classical* apparatus yields a specific measurement result.  While possibly resolving my first conundrum, it immediately raised three related issues:  why is it necessary to revert to classical physics; when and where does this classical measurement occur, i.e., where is the quantum/classical divide; and how does one describe the physical interaction across the divide.  Of course, these are questions that had vexed physicists and philosophers since the beginning of quantum mechanics but I undoubtedly thought that I had conjured them myself.  It's interesting that the notion of state reduction (collapse of the wave function) wasn't on my list of puzzlements. (It was certainly foremost on Everett's mind.)  Perhaps, this is because none of my introductory quantum texts mentioned it.  In addition, as an experimentalist, I always considered the measurement to be the end of the process.  If one wanted to continue after that, then a new experiment had to be effected.  I'll return to this topic in Section 6.

DeWitt's solution, for which he credits Everett, was that quantum theory describes everything including the experimental apparatus and even the scientist who performs the experiment.  In addition, the wave function never collapses but rather a measurement (or any interaction for that matter) initiates a branching of the universe so that all the statistically possible results of any measurement occur in one of the branches.  According to DeWitt (1970),

> This universe is constantly splitting into a stupendous number of branches,
> all resulting from measurement-like interactions between its myriad of components. Moreover, every quantum transition taking place on every star, in every galaxy, in every corner of the universe is splitting our local world on earth into myriads of copies of itself.

This was DeWitt's many worlds interpretation, although he didn't use that term at the time. He and Neill Graham edited a volume, *The Many Worlds Interpretation of Quantum Mechanics* (DeWitt and Graham 1973), which consisted of reprints of papers by Everett, DeWitt, Graham, and others. To be sure, Everett did not describe his interpretation in terms of branching worlds. Finally, a still disputed claim of Everett's and DeWitt's was that the probabilistic predictions of standard quantum theory arose naturally from the formalism with no necessity of postulating a statistical interpretation (i.e., the Born rule). Rather, "The mathematical formalism of the quantum theory is capable of yielding its own interpretation" (DeWitt 1970). The latter claim involves disputable mathematical and philosophical analyses that I won't address in this paper.

The notion of a myriad of branching universes certainly resolved my contention that "according to quantum mechanics, nothing ever happens" by claiming that "everything happens." On the other hand, at least for me, it did nothing to alleviate my worry about when and where the quantum/classical transition occurs. That problem was just replaced by the question of when and where the universe branches. The many worlds interpretation was mute on that point. Again, the branching universe concept was DeWitt's, not Everett's, and I'll return to this issue later.

In a nutshell, this was what I gleaned from DeWitt's *Physics Today* article. In the intervening years, hundreds of papers have analyzed and expanded upon Everett's thesis. While I've read some of these and will briefly mention a few in the following pages, my present essay is not intended to be a historical critique of the many worlds interpretation. The reason is that most of these studies are theoretical/philosophical in nature and invariably become involved in the ontology of quantum theory. My pragmatic, experimentalist's view is more epistemic and consequently I have little interest in pondering the nature of quantum reality. In the next section I'll present a simple toy model that captures the essence of what I believe was Everett's theory and in the following sections I'll address questions I have about such a model.

Before moving on, I should mention one other paper from my graduate student days that would eventually have a profound effect on my understanding of the foundations of quantum theory. In 1972, Henry Stapp authored a paper entitled "The Copenhagen Interpretation". To be sure, Stapp's motivation was not to express his own interpretation of quantum mechanics but to present, as accurately as possible, Niels Bohr's interpretation;[2] however, Stapp's own S-matrix interpretation of quantum mechanics accepts most of the pragmatic features of the Copenhagen interpretation (Stapp 1971). The views expressed in Stapp's paper will figure strongly in my pragmatic criticisms of the many worlds interpretation.

## 2. THE WAVE FUNCTION OF THE UNIVERSE

In January 1956 Everett submitted a long version of his thesis, *Wave Mechanics Without Probability,* to his advisor John Wheeler. It was later included in DeWitt's book (DeWitt and Graham 1973) under the title, *The Theory of the Universal Wave Function*. Wheeler carried of a version of this document to Copenhagen in order to discuss it with his good friend Neils Bohr and coworkers. It was not well received. Whether or not this situation led Wheeler to pressure Everett to significantly revise his thesis, in March 1957 Everett submitted a much shorter doctoral dissertation (about ¼ the length of the original) under the title, *On the Foundations of Quantum Mechanics*. This version was published in the July 1957 issue of *Reviews of Modern Physics* under the title *"Relative State" Formulation of Quantum Mechanics* and was accompanied by a two page paper by Wheeler (1957) assessing Everett's paper.[3] I'll come back to the importance of the notion of the *relative state* in Section 3 but in this section will concentrate on the notion of a *universal wave function*.

The very first sentence of the long version of Everett's thesis was: "We begin, as a way of entering our subject, by characterizing a particular interpretation of quantum theory which … is the most common form encountered in textbooks and university lectures on the subject." Referencing von Neumann (1955), he immediately identifies the

---

interpretation as postulating a state function description of a system and then specifies two modes of state evolution: 1) the (nonlinear) discontinuous change to an eigenstate of some operator, the *collapse of the wave function,* and 2) the unitary evolution of the state function (e.g., via Schrödinger's equation). He then points out paradoxes (of the Schrödinger's cat genre) associated with this exegesis. Recall that wave function collapse was not one of my problems with quantum mechanics. Nor was it a component of the Copenhagen interpretation. Everett acknowledged this but then dismissed Bohr's interpretation with:

> While undoubtedly safe from contradiction, due to its extreme
> conservatism, it is perhaps overcautious. We do not believe that
> the primary purpose of theoretical physics is to construct 'safe'
> theories at severe cost in the applicability of their concepts, which
> is a sterile occupation, but to make useful models which serve for
> a time and are replaced as they are outworn.

One can imagine that such a characterization of Bohr's perspective might have vexed Wheeler.

Everett's way out of the conundrum was to include the measuring apparatus (and observer for that matter) in the state function of the entire system, which then evolves only according to the unitary prescription, i.e., there is no collapse of the wave function. He then proclaims that

> Since the universal validity of the state function description is asserted,
> one can regard the state functions themselves as the fundamental entities,
> and one can even consider the state function of the whole universe. In
> this sense this theory can be called the theory of the "universal wave
> function," since all of physics is presumed to follow from this function
> alone.

In this single step, Everett rids us of the paradox of wave function collapse and of any problems associated with the inclusion of classical physics in the measurement process. After all, the observer is now part of the quantum state function. Everett then notes: "There remains, however, the question whether or not such a theory can be put into correspondence with our experience." Most of the remainder of his thesis is devoted to answering this question by invoking the "relative state" concept. I'll return to this in the next section.

Before addressing the concept of the wave function of the (entire) universe using a toy model, let me say a few words about the connection of Everett's theory with general relativity and cosmology. Because general relativity is theory of all of space and time, there is no way for an external observer to view it from the outside as is often ascribed to the Copenhagen interpretation. Therefore, a theory that includes the observer in the formal description of a system might seem like a natural fit with general relativity. In the long version of his thesis, Everett seemed unaware of such a connection or at least he made no mention of it. However his advisor, John Wheeler, certainly was, which helps explain why he took such interest in Everett's idea. As Wheeler concluded in the very last sentence of his 1957 commentary on Everett's thesis,

> Apart from Everett's concept of relative states, no self-consistent system of ideas is at hand to explain what one shall mean by quantizing a closed system like the universe of general relativity.

Presumably as a consequence of Wheeler's influence, the very first sentence of Everett's 1957 paper (the subsequent short version of his thesis) was

> The task of quantizing general relativity raises serious questions about the meaning of the present formulation and interpretation of quantum mechanics when applied to so fundamental a structure as the space-time geometry itself.

At the end of the paper he concluded that "The 'relative state' formulation…may therefore prove a fruitful framework for the quantization of general relativity." It appears that Everett got Wheeler's message loud and clear. More recently, many others have been attracted to Everett's work precisely because of this quantum mechanics-general relativity-cosmology connection. Several examples can be found in Deutsch (1985), Tipler (1986), Gell-Man & Hartle (1989), Page (1999), Aguirre & Tegmark (2011), Bousso & Suskind (2012), and Linde (2016).

Now to my toy universal wave function: Just what does one mean by a *universal wave function* and what does this have to do with DeWitt's *many worlds*? Well, such a wave function surely constitutes a quantum description of at least the $10^{80}$ particles that we know inhabit the observable universe. Such a wave function resides in a $3 \times 10^{80}$ dimensional configuration space. Each point in such a space would represent the location of all the particles in the universe, i.e., a possible configuration of the universe. Because this wave function never collapses, one is prohibited from ascribing reality to only one such point, i.e., there are a plethora of universes. Furthermore, since there are an infinite number of points in configuration space, there must be an infinite number of universes. If instead of a wave function in configuration space we use the equivalent description of vectors in an infinite dimensional Hilbert space, the same conclusion is reached. As DeWitt (1971) pointed out, this isn't quite right. Because of the uncertainty principle, not all of these universes are distinguishable. So how many of them are there? Consider the following back-of-the-envelope estimate.

Each of the $10^{80}$ particles can be located anywhere in the (observable) universe, which has a volume of approximately $10^{79} m^3$. As a smallest volume in which a particle might be localized, let's pick the Planck volume, $4 \times 10^{-105} m^3$. Surely particles cannot possibly be located more precisely than this. (You'll see in a moment that it hardly matters what smallest volume one chooses.) Thus there are about $10^{183}$ possible locations in the universe for each of the $10^{80}$ particles. If $n = 10^{183}$ and $m = 10^{80}$, then the total number of possible universes is $N = n^m$. Because quantum particles are identical one must divide by the number of configurations of $m$ particles, i.e., $m!$. By Stirling's approximation $m! \sim (m/e)^m$ to within a factor on the order of unity. Thus $N \sim \left(\frac{ne}{m}\right)^m = 10^{10^{82.0}}$. (Never did I imagine I would ever encounter a number in the googolplex range!)[4] One might worry that I didn't take into account that fermions assiduously avoid each other; however, there are so many of these cells in the universe that there is little chance that more than one particle would occupy the same Planck volume. Also, in this notation, there is little dependence on the cell size. For example, if we divided up the observable universe into cubic centimeters (1cc $\sim 10^{98}$ Planck volumes), we have $N \sim 10^{10^{80.7}}$. A plethora of universes, indeed!

So now we know roughly how many universes the universal wave function describes but just what is this universal wave function? It's hard to wrap your mind around a wave function that describes absolutely everything in the universe, including a person like you in the process of trying to wrap their mind around a universal wave function. It's difficult to imagine expressing such a wave function mathematically because that expression must also contain a description of us writing down that expression. (David Deutsch (1997) might claim that quantum computers could accomplish this). Also, how do we interpret the amplitude of the wave function? Is it a probability amplitude? It can't indicate the probability of a given universe existing because we assume that all of the universes described by the wave function exist (in some sense). Everett's answer is that the universal wave function is meaningful but only in a "relative" sense as discussed in the next section.

---

[4] The astute reader will have noticed that I considered $\sqrt{m} = 10^{40}$ to be of the order unity! Now you can see why.

Another problem is how are we ever to identify universes like ours with people like us in the nearly googolplex of universes included in the universal wave function. This is often referred to as the problem of identifying a preferred basis in Hilbert space. Clearly most of the universes will be chaotic and bear no resemblance to the one in which we live but there will still be very many with nearly identical copies of ourselves. A great deal has been written about how one might identify such bases or equivalently how universes such as ours naturally emerge. I'll briefly discuss this issue in Section 5. Everett, for his part, didn't worry about the problem and maintained that all components of all bases are "actual" universes in the same sense. He even supposed that occasionally different universes could interact with one another. Deutsch (1997) maintains that the phenomenon of quantum interference is evidence of precisely this sort of interaction.

Let us now turn to the evolution of these many worlds. I'll assume that the wave function in configuration space evolves according to the Schrödinger equation (ignoring the cosmological evolution of the universe). The Schrödinger equation is deterministic so the universal wave function is determined in both space and time, subject only to initial conditions. So we might consider an expanded notion of many worlds that consists of all possible universes at all possible times. This doesn't enlarge things much. Suppose we only consider possible universes in some small discrete time steps. Let's go with the Planck time, $\sim 10^{-43} s$, for surely nothing much happens on time scales shorter than this. The current age of the universe is roughly $10^{61}$ Planck times so that the number of possible universes since the big bang is our above estimate times $10^{61}$. As before, in terms of our googolplex sized number, this factor is of the order of unity and so doesn't increase the number of universes under consideration, relatively speaking.

Finally let's get to DeWitt's branching universes. How do they fit into our simple picture?  They don't.  At one time there are $\sim 10^{10^{82}}$ possible universes and at a time $\Delta t$ later there are the same set of universes (increased slightly because of the expansion of the universe).  Is it possible to connect any particular universe at one time step with another particular universe at the next time step?  From what I've said so far, I don't think so but I also don't see this as a problem with Everett's model.  (In Section 5, we'll see the claim that certain course grained histories can be traced through time.  Everett seemed comfortable with this notion.)  Assume that in the plethora of universes depicted by the universal wave function at a specific time, we can identify the Hilbert space vector (or vectors) that corresponds to our universe.  Then that vector not only describes us as we are now but all our memories and all of recorded history, including fossils, which we know and love.  In the next time step, there will be another googolplex of universes with histories of their own.  Certainly many of them will have precisely the properties expected for my perceived evolution through my particular universe but will any be inhabited by the same me that I am now?  The pragmatist in me would probably say "who cares".

Let me digress for a moment to share a childhood fear I developed at about the age of nine.  I became obsessed with the notion that when I fell asleep I would cease to exist and in the morning an entirely different person would awake, albeit a person with precisely my memories.  (I was also terribly frightened by the 1956 movie, *The Invasion of the Body Snatchers*, which may have had something to do with the former fear.[5])  I would hope that I got over this fear by embracing the pragmatist's "who cares" attitude but I probably just got tired of being afraid.

If one cannot trace a path though time from one version of the universe to the next, then how is one to make sense of DeWitt's branching universes?  Everett didn't seem to think this was an important question.  In fact, he never made any reference to branching universes.  In his short thesis he did refer to "observer state branches".  This brings us to the next topic, the *relativity of states*.

3.  RELATIVE STATES

---

[5] I leave it to the (older) reader as to whether that film was an allegory or satire of the fear of communism and associated McCarthyism that gripped the US in the 1950's.

After introducing the concept of the universal wave function, Everett declares in the long version of his thesis [italics are Everett's]

> *The present thesis is devoted to showing that this concept of a universal wave mechanics, together with the necessary correlation machinery for its interpretation, forms a logically self consistent description of a universe in which several observers are at work.*

What follows is a very formal and detailed analysis to establish the logical self-consistency of his approach. Perhaps the most important concept that Everett introduced in this process was the notion of a relative state, which appeared in the title of the published version of his shortened PhD thesis, *"Relative State" Formulation of Quantum Mechanics*. In the long version of his thesis, Everett (1956) puts it like this:

> There does not, in general, exist anything like a single state for one subsystem of a composite system. That is, subsystems do not possess states independent of the states of the remainder of the system, so that the subsystem states are generally correlated. One can arbitrarily choose a state for one subsystem, and be led to the relative state for the other subsystem. Thus we are faced with a fundamental relativity of states, which is implied by the formalism of composite systems. It is meaningless to ask the absolute state of a subsystem - one can only ask the state relative to a given state of the remainder of the system.

The analysis that precedes this conclusion is based on standard quantum mechanics as presented in von Neumann's, *Mathematical Foundations of Quantum Mechanics* (1955), to which Everett liberally refers. Although the analysis is extensive, as is required to establish "logical consistency", the notion can be expressed quite simply. When an observer performs an experiment on an object, the observer part of the universal wave function becomes entangled with the object part of the wave function, an entanglement represented by a superposition of possible object-observer states. Consider for example, the measurement of the z-component of the spin, $S_z$, of a spin ½ particle with a Stern-Gerlach apparatus. Then the $S_z = +\frac{1}{2}$ part of the particle wave function becomes entangled with the part of the observer wave function that sees the particle as spin up and the $S_z = -\frac{1}{2}$ part of the particle wave function becomes entangled with the part of the observer wave function that sees the particle as spin down. If we now choose to consider only the part of the wave function for which the observer sees the particle as spin up, then the "relative state" of the particle is the wave function corresponding to $S_z = +\frac{1}{2}$. According to Everett, there is no absolute particle state in isolation of the observer but rather only the entangled superposition of both outcomes[6]. Nevertheless, one can always specify the particle state that is *relative* to the state of an observer that sees a particular outcome.

Nowhere in this analysis does one assume the wave function of the particle collapses to a particular spin eigenstate. The evolution of the object-observer wave function is according to the unitary Schrödinger equation. However, if we limit our discussion to a particular outcome observed by the experimenter, then the relative state of the particle is precisely the same as if the observation caused the wave function to collapse in the usual sense. That is, Everett's interpretation is completely consistent with the "textbook interpretation of quantum mechanics" (as characterized by Everett).

So after the experiment has been performed and has determined that the particle is in a spin up state, what is Everett's interpretation of the spin down part of the superposition? According to Everett (1956)

> It is therefore improper to attribute any less validity or "reality" to any element of a superposition than any other element, due to this ever present possibility of obtaining interference effects between the elements. All elements of a superposition must be regarded as simultaneously existing.

That is, one must not toss out the other terms in the superposition because, no matter how extraordinary difficult it might be, it is always possible in principle to design a subsequent experiment that is sensitive to the other terms. It is in this sense that Everett considers all terms of the superposition equally "real". Everett only uses the term "branches" with regard to "trajectories" of observers. He mentions this only once in the long version of his thesis but is more detailed in his "relative state" paper (Everett 1957):

> We thus arrive at the following picture: Throughout all of a sequence of observation processes there is only one physical system representing the observer, yet there is no single unique state of the observer (which follows from the representations of interacting systems). Nevertheless, there is a representation in terms of a *superposition*, each element of which contains a definite observer state and a corresponding system state. Thus with each succeeding observation (or interaction), the observer state "branches" into a number of different states. Each branch represents a different outcome of the measurement and the *corresponding* eigenstate for the object-system state. All branches exist simultaneously in the superposition after any given sequence of observations.

---

[6] This may strike one as reminiscent of Bohr's contention that a particular property of a quantum state cannot be specified in the absence of the apparatus used to measure it.

It is interesting that the very first time he mentions "branches" in his thesis he is referring not to the branching of the universe nor even to the branching of an observer but rather to the branching of observer *states*.  He clarified his position in a footnote on the same page:

In reply to a preprint of this article some correspondents have raised the question of the "transition from possible to actual", arguing that in "reality" there is - as our experience testifies - no such splitting of observer states, so that only one branch can ever actually exist. Since this point may occur to other readers the following is offered in explanation.  The whole issue of the transition from "possible" to "actual" is taken care of in the theory in a very simple way—there is no such transition, nor is such a transition necessary for the theory to be in accord with our experience. From the viewpoint of the theory all elements of a superposition (all "branches") are "actual", none any more "real" than the rest.  It is unnecessary to suppose that all but one are somehow destroyed, since all the separate elements of a superposition obey the wave equation with complete indifference to the presence or absence ("actuality" or not) of any other elements. This total lack of effect of one branch on another also implies that no observer will ever be aware of any "splitting" process.[7]

It's a little hard to pin down what Everett means by "real".  In the long version of his thesis, with regard to physically separated, entangled systems, he says

From the present viewpoint all elements of this superposition are equally "real".  Only the observer state has changed, so as to become correlated with the state of the near system and hence naturally with that of the remote system also. The mouse does not affect the universe - only the mouse [observer] is affected.

Perhaps the most straightforward statement about reality appears in an unpublished draft,

---

[7] Given his previous statement of the "possible" interference of the different branches, this later statement should obviously be taken as "total lack of effect" under normal circumstances.

"Probability and Wave Mechanics", that he gave to Wheeler in late 1956 (Byrne 2010, p 140):

> The physical 'reality' is assumed to be the wave function of the whole universe itself. By properly interpreting the internal correlations in this wave function it is possible to explain the appearance of the macroscopic world to us, as well as the apparent probabilistic aspects.

Perhaps this is why Everett wasn't willing to say more about the "reality" of the other branches. For him, the universal wave function is *the* reality, that is, the only reality necessary so that the "theory can be put into correspondence with our experience." DeWitt (1970) on the other hand was willing to consider each isolated branch as a separate, real universe.

For example, in his 1970 *Physics Today* paper, DeWitt described the Schrödinger's cat experiment as follows:

> The animal trapped in a room together with a Geiger counter and a hammer, which, upon discharge of the counter, smashes a flask of prussic acid. The counter contains a trace of radioactive material - just enough that in one hour there is a 50% chance one of the nuclei will decay and therefore an equal chance the cat will be poisoned. At the end of the hour the total wave function for the system will have a form in which the living cat and the dead cat are mixed in equal portions. Schrodinger felt that the wave mechanics that led to this paradox presented an unacceptable description of reality. However, Everett, Wheeler and Graham's interpretation of quantum mechanics pictures the cats as inhabiting two simultaneous, noninteracting, but equally real worlds.

Later, he and Graham describe Everett's theory with (DeWitt & Graham 1973):

> In 1957, in his Princeton doctoral dissertation, Hugh Everett, III, proposed a new interpretation of quantum mechanics that denies the existence of a separate classical realm and asserts that it makes sense to talk about a state vector for the whole universe. This state vector never collapses and hence reality as a whole is rigorously deterministic. This reality, which is described jointly by the dynamical variables and the state vector, is not the reality we customarily think of, but is a reality composed of many worlds. By virtue of the temporal development of the dynamical variables the state vector decomposes naturally into orthogonal vectors, reflecting a continual splitting of the universe into a multitude of mutually unobservable but equally real worlds, in each of which every good measurement has yielded a definite result and in most of which the familiar statistical quantum laws hold.

It seems clear that DeWitt and Graham consider that the multitude of branching worlds are "real" in the ordinary sense of the word. In this sense, their Many Worlds perspective certainly departs from Everett's intent.

In a 1976 philosophy paper on the interpretation of quantum mechanics, Lévy-Leblond offers critical comments on the many worlds interpretation and compared it to his understanding of Everett's theory.

Since, with each successive measurement, this state-vector "splits" into a superposition of several "branches", it is said to describe "many universes", one for each of these branches. Where the Copenhagen interpretation would arbitrarily choose "one world" by cutting off all "branches" of the state-vector except one (presumably the one we think we sit upon), one should accept the simultaneous existence of the "many worlds" corresponding to all possible outcomes of the measurement. Now, my criticism here is exactly symmetrical of the one I directed against the orthodox position: the "many worlds" idea again is a left-over of classical conceptions. The coexisting branches here, as the unique surviving one in the Copenhagen point of view, can only be related to "worlds" described by classical physics. The difference is that, instead of interpreting the quantum "plus" as a classical "or", De Witt *et al.* interpret it as a classical "and". To me, the deep meaning of Everett's ideas is not the coexistence of many worlds, but on the contrary, the existence of a single quantum one. The main drawback of the "many-worlds" terminology is that it leads one to ask the question of "what branch we are on", since it certainly looks as if our consciousness definitely belonged to only one world at a time: But this question only makes sense from a classical point of view, once more. It becomes entirely irrelevant as soon as one commits oneself to a consistent quantum view.

In a letter to Lévy-Leblond (Barrett 2011), Everett indicated that he quite agreed with Lévy-Leblond's argument and emphasized that the many worlds terminology was not his.[8] I'm sympathetic with this view. In fact, I suspect that most of the conundrums and paradoxes associated with quantum mechanics arise from trying to impose classical

_____________________

[8] However, in the same letter he expressed his gratitude to DeWitt: "[The] Many-Worlds Interpretation etc. ...is not my title as I was pleased to have the paper published in any form anyone chose to do it in!... Far be it for me to look a gift Boswellian writer in the mouth!"

views on what are essentially quantum phenomena.[9]

## 4.  THE CORRESPONDENCE OF QUANTUM THEORY WITH EXPERIENCE

The first part of Everett's (long) thesis established the self-consistency of his theory of the universal wave function. He then undertakes to address "the question whether or not such a theory can be put into correspondence with our experience."  This endeavor is undertaken in a section entitled "OBSERVATION".  In it, he models a measurement as an interaction between the object and observer that results in a superposition of entangled object-observer states and then demonstrates that the results of a sequence of observations involving one or more observers reproduces the results predicted by standard quantum mechanics for *almost all* observers.[10]  In the limit of an infinite number of observations, the number of observers who find results inconsistent with standard quantum mechanics comprise a set of measure zero.  Everett accomplished this by introducing "a general scheme for assigning a measure to the elements of a superposition of orthogonal states" and concludes "that the only choice of measure consistent with our additivity requirement is the square amplitude measure."[11]  This was the source of his claim that the statistical interpretation of quantum mechanics follows directly from the formalism, with no need to postulate a "Born rule".  Everett's conclusion has been criticized by many but, as I mentioned before, it is irrelevant to my deliberations and so I won't comment on it further.

What Everett did not do was to indicate how the unitary laws of quantum mechanics *lead* to the branching of observer states that are consistent with what we perceive as our "real" world.[12]  He didn't seem to be interested in such details.  All he endeavored to do was to show that our world was describable in the context of a universal wave function.  If so, then his theory was empirically faithful.  As Barrett (2016) put it,

> So, for Everett, a theory was *empirically faithful* and hence empirically acceptable if there was a homomorphism between its model and the world as experienced. What this amounted to here was that pure wave mechanics is empirically faithful if one can find observers' experiences appropriately associated with modeled observers in the model of the theory. In short, Everett took pure wave mechanics to be empirically faithful because one could find quantum mechanical experience in the model as relative memory records associated with relative modeled observers.

---

[9] I made this point in a similarly titled paper about Bell's theorem and quantum non-locality. (Boughn 2017)

[10] That there will be some "universes" in which the Born rule is invalid is not particularly problematic just as within standard quantum mechanics there is a finite probability that some observers will find the results of a set of measurements to be extremely (statistically) unlikely.

[11] Everett points out that "The situation here is fully analogous to that of classical statistical mechanics, where one puts a [Lebesgue] measure on trajectories of systems in the phase space by placing a measure on the phase space itself."

However, Barrett continues,

> One might, of course, want more than empirical faithfulness from a satisfactory formulation of quantum mechanics. In keeping with his view that pure wave mechanics is quantum mechanics without probabilities, Everett simply conceded that every relative state under every decomposition of the absolute state in fact obtains. The resulting problem, one might feel, is that empirical faithfulness, in Everett's sense at least, is a relatively weak form of empirical adequacy. This can be seen by considering how one should understand the very notion of having a differential expectations when every physically possible measurement result is in fact fully realized in the model of the theory… That Everett's notion of empirical faithfulness is a relatively weak version of empirical adequacy, then, is exhibited in what pure wave mechanics, being empirically faithful, does not explain. In particular, it does not explain what it is about the physical world that makes it appropriate to expect one's relative sequence of records to be typical in the normsquared-amplitude sense, or any other sense. In short, while one can get subjective expectations for future experience by stipulation, the theory itself does not describe a physical world where such expectations might be understood as expectations concerning what will in fact occur.

I wonder if Everett would have appreciated the irony of having his theory characterized as being "relatively weak" vis-à-vis empirical adequacy after having leveled an analogous criticism of Bohr's interpretation (see Section 2).

---

[12] Recall my exhortation: "Another problem (for me) is how are we ever to identify universes like ours with people like us in the nearly googolplex of universes included in the universal wave function."

5.  EMERGENCE OF THE "CLASSICAL" WORLD

It is the consensus among physicists today that all forms of matter and radiation, including experimental apparatus and observers, are in principle governed by the laws of quantum mechanics. Bohr and Heisenberg were certainly comfortable with this view, contrary to how some people characterize the Copenhagen interpretation.  (I'll comment more about this in the next section.)  Consequently, there has been considerable effort to understand the "classical" world we experience in terms of its quantum mechanical description.  Progress has been made in identifying preferred Hilbert space bases that can be identified with the world we experience, which would certainly correct one of the primary shortcomings of Everett's thesis and that is how our classical world emerges from the universal wave function.  In fact, Everett's theory served as inspiration for some of this work.

Perhaps the most successful of these forays has involved the phenomenon of decoherence that was advanced by Zeh in the early 1970s and Zurek in the early 1980s.[13] As we described it elsewhere (Boughn & Reginatto 2013)

> Decoherence theory is neither new physics nor a new interpretation of quantum mechanics; although, it is certainly relevant to questions of interpretation. In decoherence theory, the measuring apparatus and the environment with which it inevitably interacts are both treated as purely quantum mechanical systems. As a consequence of the interactions of the quantum system of interest with the measuring apparatus and it with its immediate environment, the three become entangled, i.e., strongly correlated with each other. All, or at least most, of the environmental quantum degrees of freedom are not observable (certainly, not observed) and, therefore, must be summed over to achieve a reduced state of the system plus apparatus. The net effect of the enormous number of environmental degrees of freedom is that off-diagonal terms of the reduced density matrix rapidly vanish, i.e., coherence between the different eigenstates of the system/apparatus is lost. Thus, decoherence

---

[13] Zurek credits Everett for inspiring him to pursue this phenomenon. (Byrne 2010, p. 366)

theory demonstrates why it is that quantum coherence is seldom, if ever, observed at the classical (macroscopic) level.

While much of the work on decoherence has not been directly related to the "many worlds" aspect of Everett's thesis, it is directly related to the notion of a universal wave function in the context of cosmology. For example, Gell-Mann and Hartle (1990) were motivated by developing "a quantum-mechanical framework for the universe as a whole…for describing the ultimate origin in quantum cosmology of the 'quasiclassical domain' of familiar experience and for characterizing the process of measurement." The decoherence histories approach that they promote involves the course graining of decoherent trajectories in Hilbert space together with a probability measure for these trajectories. In addition, it is possible for some course grained "universes" to have consistent histories over time in just the way we see our universe as normally evolving. Therefore, the decoherent histories formalism alleviates both my discomfort with there being a huge number of copies of myself with virtually identical experiences and my childhood fear of ceasing to exist every night when I fell asleep (or more drastically at every time step of the universal wave function). While many such analyses do not directly reference the "many worlds" concept, like Everett's, their purpose was to develop an approach that "does not depend on an assumed separation of classical and quantum domains, on notions of measurement, or on collapse of the wave function." (Halliwell 1995).

While I find these sorts of analyses interesting, even compelling, I long ago came to doubt that there was any need to resolve the quantum/classical dilemma. For me, Schrödinger's cat poses no paradox. I certainly assumed that the cat could be treated quantum mechanically, in which case the quantum state would consist of the superposition of a live and dead cat. Like Everett, I also assumed that, in principle, one could perform an experiment that would reveal the presence of both the live and dead states. On the other hand, the difficulty of performing such an experiment would render bringing the dead back to life trivial by comparison. Also, the sorts of analyses referred to above deal only in generalities. They don't even approach specifying a detailed macroscopic quantum state or an associated Hamiltonian. Finally, even if I were to find the arguments compelling, I would characterize them as simply demonstrating the *consistency* of the quantum mechanical microscopic world with the classical macroscopic world rather than demonstrating how classical physics *emerges* from quantum mechanics. I suppose this isn't a very important distinction. In any case, many worlds or not, I will level some criticism of these cosmological considerations in the next section.

## 6. CHALLENGING *MANY WORLDS*

From my point of view, there are four problems of "conventional" quantum mechanics that Everett chose to confront in his thesis: 1) the non-unitary collapse of the wave function; 2) the Copenhagen interpretation's reliance on the classical description of measurements; 3) the probabilistic assumption (Born rule) that was necessary to interpret quantum formalism; and 4) the problem of "quantizing general relativity" and the "serious questions about the meaning of the [then] present formulation and interpretation of quantum mechanics when applied to so fundamental a structure as the space-time geometry itself" (Everett 1957)[14]. Everett felt he successfully addressed problem 3) by demonstrating that the probabilistic assertions of quantum mechanics were subjective appearances of observers that follow directly from the quantum formalism. As I mentioned before, this conclusion has been widely criticized as well as expanded upon by others and I won't elaborate on it. From my pragmatic perspective, the Born rule is just fine.

Everett's vehicle for confronting all these problems is "the universal wave function" so it is here that I'll begin my challenge. As I mentioned before, he didn't seem particularly interested in discussing the "reality" of what DeWitt referred to as the myriad of branches of the universe. For Everett, these branches were peripheral to his contention that "the physical 'reality' is assumed to be the wave function of the whole universe itself." (Bryne 2010, p. 140) Even though these were Everett's words, it is not entirely clear just what he meant by "reality". Perhaps it's telling that in both the long and short versions of his thesis the words "real" and "reality" only appear within quotation marks. Maybe he was trying to tell us that he didn't mean the sort of ontic reality that we (and DeWitt) all know and love. Another indication of Everett's take on realism appeared in the second appendix to the preliminary, long version of his thesis. The title of the appendix was "Remarks on the Role of Theoretical Physics" and in it he notes:

> The essential point of a theory, then, is that it is a *mathematical model*… However, when a theory is highly successful and becomes firmly established, the model tends to become identified with "reality" itself, and the model nature of the theory becomes obscured. The rise of classical physics offers an excellent example of this process. The constructs of classical physics are just as much fictions of our own minds as those of any other theory we simply have a great deal more confidence in them. It must be deemed a mistake, therefore, to attribute any more "reality" here than elsewhere… Once we have granted that any physical theory is essentially only a model for the world of experience, we must renounce all hope of finding anything like "the correct theory."

---

[14] Recall that the latter problem did not appear in the preliminary, long version of Everett's thesis but, presumably at Wheeler's request, the final short version began by emphasizing it. This problem is considered by many to be of paramount importance to quantum cosmology.

Everett seems to be telling us that the notion of a universal wave function is actually just a "fiction of our own minds".   According to Barrett (2016), Everett was "an operational realist",

> Rather than take the branches determined by a physically preferred basis or those determined by, or roughly determined by, some decoherence condition to determine which physically possible worlds were real, he took every branch in any basis to have observational consequences and hence to be real in his operational sense. Given how he understood branches and their role in determining the empirical faithfulness of the theory, Everett never had to say anything concerning how a particular physically preferred basis is selected because none was required.

Nevertheless, much of the discussion of the interpretation of quantum mechanical wave functions, especially in the philosophical literature has focused on their ontological nature.

In order to comprehend the wave function of the universe, it necessary to address the question of what we mean by the term "wave function".  By "mean" I'm referring to the empirical meaning that any experimentalist like me needs to know.  It is here that Stapp's 1972 paper on the Copenhagen interpretation informs us.  In Stapp's practical account of quantum theory, a system to be measured is first prepared according to a set of specifications, $A$, which are then transcribed into a wave function $\Psi_A(x)$, where $x$ are the degrees of freedom of the system. The specifications $A$ are "couched in a language that is meaningful to an engineer or laboratory technician", i.e., not in the language of quantum (or even classical) formalism. Likewise, $B$ are a set of specifications of the subsequent measurement and its possible results. These are transcribed into another wave function $\Psi_B(y)$, where $y$ are the degrees of freedom of the measured system. How are the mappings of $A$ and $B$ to $\Psi_A(x)$ and $\Psi_B(y)$ effected? According to Stapp,

> …no one has yet made a qualitatively accurate theoretical description of a measuring device. Thus what experimentalists do, in practice, is to *calibrate* their devices…[then] with plausible assumptions…it is possible to build up a catalog of correspondences between what experimentalists do and see, and the wave functions of the prepared and measured systems. It is this body of accumulated empirical knowledge that bridges the gap between the operational specifications $A$ and $B$ and their mathematical images $\Psi_A$ and $\Psi_B$. Next a transition function $U(x; y)$ is constructed in accordance with certain theoretical rules…the 'transition amplitude' $\langle A|B \rangle \equiv \int \Psi_A(x)\, U(x; y) \Psi_B^*\, dx dy$ is computed. The predicted probability that a measurement performed in the manner specified by $B$ will yield a result specified by $B$, if the preparation is performed in the manner specified by $A$, is given by $P(A, B) = |\langle A|B \rangle|^2$.

Of course, our use of quantum wave functions is not limited to the experimentalist's laboratory. One often assigns a wave function to systems and measurements that have been prepared *naturally*. Consider, for example, the case of the emission of 21 cm radiation from hydrogen in the interstellar medium. In that case one specifies the intial upper ground state wave function, the subsequent lower ground state wave function, and the interaction that results in the emission of radiation. In this case, the interstellar environment takes the place of the experimentalist's laboratory. Nevertheless, the conditions of this environment provide the operational specifications that the astronomer must translate into wave functions. Also, it is unlikely that such translations could ever be made without the laboratory experiments that informed us about quantum phenomenon.

What's my point here? It is that the quantum mechanical wave function is a theoretical construct that we invented to deal with our observations of physical phenomena. As such, it seems reasonable that we derive our understanding of it according to how we use the concept. So, in what sense is it reasonable to talk about the wave function of the entire universe? Stapp's (and Bohr's) pragmatic account of wave functions is intimately tied to state preparation and measurement, both of which are described in terms of operational specifications that lie wholly outside the formalism of quantum mechanics. Prior to the preparation of a system the wave function is not even defined and after it has been measured the wave function ceases to have a referent. To extrapolate the notion of the wave function to the entire universe is an enormous leap that necessitates dropping all references to the operational specifications that gave wave functions their meanings in the first place. One might be led to such an extrapolation by ascribing an ontic reality to the notion of wave function (despite its epistemic transitory nature). This is especially tempting when presented with some of the incredible successes of quantum theory. I was certainly enticed to do so when I first learned of the twelve decimal place agreement of the quantum electrodynamic prediction with the measured value of the g-factor of the electron.[15] It's interesting that this is precisely what Everett cautioned us against in the appendix to the long version of his thesis, and yet he still embraced the notion of a universal wave function. To accept the notion of a universal wave function requires a careful treatment of the meaning of the concept. Simply transferring the construct from its pragmatic atomic physics origins is, for me, not good enough. Now to the specific problems that Everett confronted.

On the very first page of the preliminary version of his thesis Everett presents what he terms as "the most common form [of quantum mechanics] encountered in textbooks and university lectures on the subject" and identifies the problem of "The discontinuous change brought about by the observation", i.e., the collapse of the wave function. As I've already mentioned, wave function collapse wasn't a concept that I encountered in my introductory quantum mechanics courses. So just how did this concept enter the quantum lexicon? Some credit (blame?) Heisenberg from comments in his seminal paper on the uncertainty principle (Heisenberg 1927). I can find only two relevant comments. He notes that as a consequence of observing an atomic system "we will find the atom has jumped from the $n^{th}$ [superposition] state to the $m^{th}$ state with a

---

[15] When I was a graduate student it was considered prescient that the acronym for quantum electrodynamics (QED) was the same as for the Latin phrase *quod erat demonstrandum*.

probability…" and in a discussion of an electron wave packet he comments, "Thus every position determination *reduces* [my italics] the wavepacket…" Perhaps, the first careful discussion of wave function reduction appeared in von Neumann's 1932 revolutionary text *The Mathematical Foundations of Quantum Mechanics* (von Neumann 1955) in which he gave a quantum mechanical description of measurements. He specified two processes in quantum mechanics: the discontinuous change of states that occurs as a result of a measurement; and the unitary evolution of quantum states (e.g., the Schrödinger equation). In 1935, Dirac was more explicit in his equally revered text, *The Principles of Quantum Mechanics,* (Myrvold 2017)

> When we measure a real dynamical variable $\xi$, the disturbance involved in the act of measurement causes a jump in the state of the dynamical system. From physical continuity, if we make a second measurement of the same dynamical variable $\xi$ immediately after the first, the result of the second measurement must be the same as that of the first. Thus after the first measurement has been made, there is no indeterminacy in the result of the second. Hence, after the first measurement has been made, the system is in an eigenstate of the dynamical variable $\xi$, the eigenvalue it belongs to being equal to the result of the first measurement. This conclusion must still hold if the second measurement is not actually made. In this way we see that a measurement always causes the system to jump into an eigenstate of the dynamical variable that is being measured, the eigenvalue this eigenstate belongs to being equal to the result.

This seems like a reasonable conclusion; so what is my problem with it? For one thing, measurements rarely proceed in such a fashion. After a measurement, the object that was given a wave function description is usually nowhere to be found. Even a relatively simple experiment as provided by a Stern-Gerlach apparatus usually ends with a polarized atom striking a photographic emulsion (or the modern equivalent) after which there is no conceivable way to measure its state again. To be sure, in special cases the quantum system can be measured to be in a particular eigenstate and afterwards can be found in that same state. In those cases it seems as if the wave function might actually have collapsed.

Recall that Everett maintained such a discontinuous transition never occurs. His universal wave function continues evolving according to the Schrödinger equation. The apparent collapse is simply in the mind of a particular state of the observer. DeWitt might replace "particular state of the observer" with "particular observer" with the implication that there are many such observers (many worlds). Bohr would agree that there is no such thing as the collapse of the wave function. So how did he explain the "apparent" collapse of the wave function? The Copenhagen interpretation's resolution of the problem (according to Stapp 1972) is as straightforward as Everett's. In the case that the quantum system is disrupted upon the completion of a measurement, the pre-measurement wave function simply ceases to have meaning and there is nothing that requires the notion of wave function collapse. In the special cases mentioned above, the measurement does not disrupt the quantum system but leaves it in a state with a well-defined wave function that can be used to predict a subsequent measurement. This is the only time that the notion of wave function collapse might make sense. However, in this case, the Copenhagen interpretation would simply point out that the original specifications of the quantum system must be replaced by new ones that include the specifications of the results of the measurement. "Then the original wave function will be naturally replaced by a new one, just as it would be in classical statistical theory." (Stapp 1972) Some physicists prefer to claim, euphemistically, that wave function collapse happens in the mind of the observer but not for any physical reason. For example Hartle, who took Everett's ideas seriously, noted (Hartle 1968)

> The "reduction of the wave packet" does take place in the consciousness of the observer, not because of any unique physical process which takes place there, but only because the state is a construct of the observer and not an objective property of the physical system.

The second problem with conventional quantum mechanics that Everett sought to solve was the Copenhagen's interpretations requirement that the act of measurement or observation must be described in terms of classical physics. As Everett (1956) put it,

> Another objectionable feature [in addition to its "extreme conservatism" referred to in Section 2] of this position is its strong reliance upon the classical level from the outset, which precludes any possibility of explaining this level on the basis of an underlying quantum theory. (The deduction of classical phenomena from quantum theory is impossible simply because no meaningful statements can be made without pre-existing classical apparatus to serve as a reference frame.) This interpretation suffers from the dualism of adhering to a "reality" concept (i.e., the possibility of objective description) on the classical level but renouncing the same in the quantum domain.

Many others have worried about the quantum/classical divide and it is usually considered to be one of the major conundrums of the *measurement problem* in quantum mechanics. The dilemma is predicated upon the supposition that experiments must be, or inevitably are, described by classical physics. So why didn't Bohr seem to worry about the unseemly merger of classical and quantum formalism? I suspect that the answer is because he didn't actually consider that any such merger is required. While Bohr endeavored to be extremely careful in expressing his ideas, his prose is often obscure. However, consider his following brief description of a measurement (Bohr 1963, p. 3):

> The decisive point is to recognize that the description of the experimental arrangement and the recordings of observations must be given in plain language, suitably refined by the usual terminology. This is a simple logical demand, since by the word 'experiment' we can only mean a procedure regarding which we are able to communicate to others what we have done and what we have learnt.

Nowhere in this description does he refer to classical physics. Stapp (1972) chose to emphasize this pragmatic view by using the word specifications , i.e.,

> Specifcations are what architects and builders, and mechanics and machinists, use to communicate to one another conditions on the concrete social realities or actualities that bind their lives together. It is hard to think of a theoretical concept that could have a more objective meaning. Specifications are described in technical jargon that is an extension of everyday language. This language may incorporate concepts from classical physics. But this fact in no way implies that these concepts are valid beyond the realm in which they are used by technicians.

The point is that descriptions of experiments are invariably given in terms of operational prescriptions or specifications that can be communicated to the technicians, engineers, and the physics community at large. Are these operational prescriptions part and parcel of classical theory? Are they couched in terms of point particles, rigid solid bodies, Newton's laws or Hamilton-Jacobi theory? Of course not. They are part of Bohr's "procedure regarding which we are able to communicate to others what we have done and what we have learnt." Camilleri and Schlosshauer (2015) point out, "Bohr's doctrine of classical concepts is not primarily an interpretation of quantum mechanics (although it certainly bears on it), but rather is an attempt by Bohr to elaborate an epistemology of experiment." So it seems to me that Everett's (and others') assertion that Bohr insisted on merging a quantum system with a classical measuring apparatus is a "straw man" that has little to do with Bohr's interpretation of quantum mechanics. In fact, one can find several passages in Bohr's writings where he claims that the physics of experimental apparatus is certainly describable by the formalism of quantum mechanics (Bohr 2010, Camilleri & Schlosshauer 2015). But that description has little to do with the "epistemology of experiment".

In some sense, classical physics has precisely the same problem in that experiments are not described by the formalism of the theory but by the same operational specifications to which Stapp refers. (Boughn and Reginatto 2018a) This point has certainly not gone unnoticed by those trying to come to grips with the quantum/classical divide. At a 1962 conference on the foundations of quantum mechanics, Wendell Furry explained [Furry 1962]

> So that in quantum theory we have something not really worse than
> we had in classical theory. In both theories you don't say what you do
> when you make a measurement, what the process is. But in quantum
> theory we have our attention focused on this situation. And we do
> become uncomfortable about it, because we have to talk about the effects
> of the measurement on the systems....I am asking for something that the
> formalism doesn't contain, finally when you describe a measurement. Now,
> classical theory doesn't contain any description of measurement. It doesn't
> contain anywhere near as much theory of measurement as we have here
> [in quantum mechanics]. There is a gap in the quantum mechanical theory
> of measurement. In classical theory there is practically no theory of
> measurement at all, as far as I know.

At that same conference Eugene Wigner put it like this [Wigner 1962]

> Now, how does the experimentalist know that this apparatus will measure
> for him the position? "Oh", you say, "he observed that apparatus. He looked
> at it." Well that means that he carried out a measurement on it. How did he
> know that the apparatus with which he carried out that measurement will tell
> him the properties of the apparatus? Fundamentally, this is again a chain
> which has no beginning. And at the end we have to say, "We learned that as
> children how to judge what is around us." And there is no way to do this
> scientifically. The fact that in quantum mechanics we try to analyze the
> measurement process only brought this home to us that much sharply.

Because physicists have long since become comfortable with the relation between theory and measurement in classical physics, perhaps the quantum case shouldn't be viewed as particularly problematic, in which case Everett's universal wave function has nothing to offer on this point.

Finally, let me say a few words on the necessity of a universal wave function in the treatment of quantum cosmology. Because I have little knowledge of this topic, I'll have much less to say. The connection of Everett's interpretation and quantum cosmology is clear. A quantum cosmological model of the universe clearly requires a universal wave function. As Everett noted in the short, final version of his thesis:

> How is one to apply the conventional formulation of quantum mechanics to the space-time geometry itself? The issue becomes especially acute in the case of a closed universe. There is no place to stand outside the system to observe it. There is nothing outside it to produce transitions from one state to another.

Here again is the reference to the "classical" observer that the Cophenhagen interpretation considers to be external to the system. The implication is that any observer must be included in the universal wave function just as Everett's theory demands. Wheeler (1957) made a similar point in his commentary of Everett's thesis and since then many physicists with an interest in quantum cosmology have made the point that a closed system like the universe is incompatible with the notion of an external observer.[16] Some, but not all, of these reference Everett's thesis. I don't find this sort of argument at all convincing. I suspect that it, in part, arises from the mistaken characterization of Bohr's observer and measurement apparatus as elements of classical theory. As I (and Stapp) argued above, the operational specifications describing observations are external to the formalism of any theory, including quantum and classical mechanics. Everett seems to acknowledge this in the aforementioned second appendix to his long thesis. In it he explains

> Every theory can be divided into two separate parts, the formal part, and the interpretive part. The formal part consists of a purely logico-mathematical structure, i.e., a collection of symbols together with rules for their manipulation, while the interpretive part consists of a set of "associations," which are rules which put some of the elements of the formal part into correspondence with the perceived world. The essential point of a theory, then, is that it is a *mathematical model*, together with an *isomorphism* between the model and the world of experience (i.e., the sense perceptions of the individual, or the "real world" – depending upon one's choice of epistemology).

---

[16] Here is a sample of work that have made similar points: Aguirre & Tegmark 2011; Bousso & Suskind 2012; Castagnino, et. al 2017; Gell-Mann & Hartle 1990: Halliwell 1995; Hartle & Hawking 1983;

What Everett (and others) fail to understand is that Bohr's description of observers and experiments is in terms of the world of experience (sense perceptions) and not in terms of the formalism of classical physics. It seems to me that this is true of the descriptions of observations in any field of science.

Of course, one might insist that the theory of the entire universe is different than other theories and needs no "correspondence with the perceived world". This is hard for me to swallow. Cosmological theories are still models and are certainly distinct from us and our perceptions of the world. On the other hand, I have no problem with postulating a wave function of the early universe that might help us comprehend cosmology. But we must be very clear how we propose to use such a wave function and especially how we can find experimental/observational corroboration of the theory. Such a quantum cosmological wave function might be very similar to our current understanding of quantum states or it might not be. Finally, classical cosmology has provided a very successful description of the large-scale structure of the universe without ever worrying that observers aren't included in that formalism.

## 7. FINAL REMARKS AND A CONFESSION

Let me begin with the confession. When I decided to learn more about the many worlds interpretation I had already assumed I would conclude that it was a silly notion. After reading the long and short versions of Everett's thesis several times, I decided that his theory wasn't as silly as I had imagined. In fact, if I had to choose between Everett's interpretation with its universal wave function and interpretations involving the collapse of the wave function, I would go with Everett's. This is especially so because of his reluctance to ascribe to DeWitt's parallel universes. He simply postulated a universal wave function to be operationally real and stopped there. Of course, for reasons discussed above I find the idea of a universal wave function neither well defined nor necessarily useful.

I've endeavored to make a distinction between Everett's universal wave function and DeWitt's many worlds interpretation of Everett's work. For Everett, there is one, quantum universe that is characterized by a universal wave function while for DeWitt there are a "myriad of copies" of the universe. Today the term *multiverse* is more common than *many worlds*. I find it somewhat amusing that the 19[th] century American philosopher William James first used this term in an analogous discussion of the universe. I've referred to Bohr's (and my) pragmatic view of quantum mechanics. Well, James was one of the founders of philosophic pragmatism (also called radical empiricism). In 1908 he delivered a series of eight Hibbert lectures at Manchester College, Oxford, entitled *A Pluralistic Universe*. Near the end of the second lecture he characterizes the rationalist (monistic idealist) contention that one must choose between a pure (singular) universe and a pure (disconnected) multiverse. According to the rationalist there is no alternative. This distinction strikes me as precisely the difference between Everett's pure (singular) quantum universe characterized by a universal wave function and DeWitt's (disconnected) multiverse. I suspect that for them, there is no alternative to these two interpretations. James, the pragmatist, refused to choose between the two. For him and Bohr (and me), a pluralistic view of the universe is preferable.[17]

Even though Bohr dismissed Everett's interpretation and Everett clearly didn't think highly of the Copenhagen interpretation, the two certainly had something in common. Neither considered wave function collapse to be viable or even necessary. Also, they both were keenly aware that quantum mechanics was a mathematical model that needed to be put into correspondence with the perceived world. The problem was that Everett seemed not to understand that Bohr considered experiments as part of the perceived world of experience and, therefore, quite distinct from the theoretical model. Everett (and many others) interpreted Bohr's statements about classical experiments as a requirement that classical physics must be included in the formalism of quantum mechanics. Had Bohr been more kind, he might have explained to Everett the fallacy of this position.  Maybe he did try.  For his part, Everett seemed equally dismissive of the Copenhagen interpretation, which he termed "a sterile occupation".  Maybe he should have put more effort in figuring out what Bohr actually meant (even though this is no simple task).

---

[17] In fact, Stapp's (1972) correspondence with Heisenberg included a discussion of Bohr's favorable disposition towards James's philosophy.

Perhaps because of Bohr's negative impression of Everett's theory, some have argued that his universal wave function was given short shrift and, for that reason, Everett was pushed out of academia.  Cosmologist Max Tegmark, a strong supporter of Everett's interpretation, noted in a paper with Wheeler (Tegmark & Wheeler 2001), "Everett's work was largely disregarded for about two decades."  In another paper (Tegmark 2007) he opined, "Everett's discovery…took over a decade until it started getting noticed. But it was too late for Everett who left academia disillusioned."  My impression is somewhat different and so I would like to push back a bit against this conclusion.  As for Everett leaving academia disillusioned, I point out that in September 1956, six months before submitting his thesis, he left Princeton to take a job in the Pentagon's newly formed Weapons Systems Evaluation Group, a job which he seemed to quite relish (Byrne 2010).  To be sure, in 1959 at Wheeler's request he visited Bohr in Copenhagen with "disastrous results" but as far as I can ascertain he had no intensions of leaving his Washington job.  As for his work not being noticed, this also seems to be an overstatement.  In 1957 at the Chapel Hill *Conference on the Role of Gravitation in Physics* 9 (at which he was not in attendance), Everett's ideas were included in the discussion.  Even Feyman, who attended the conference, offered critical remarks on "the concept of a 'universal wave function'." (Osnaghi et al. 2009)  It was in the *Reviews of Modern Physics* collection of papers from this conference that the published version of Everett's thesis appeared.  According to Osnaghi et al. (2009),

> These were certainly not the optimum publishing conditions for Everett's work to receive the wider recognition that Wheeler had originally hoped for.  Pre-prints were nonetheless sent to many distinguished physicists, including Schrödinger, van Hove, Oppenheimer, Dyson, Yang, Wiener, Wightman, Wigner, and Margenau, besides of course Bohr and his collaborators.  The responses of DeWitt, Wiener and Margenau were quite favorable.

"Optimum" or not, such an introduction to Everett's work seems to me to be quite remarkable.  Five years later, Everett was invited to discuss his ideas at a 1962 conference at Xavier University. This was a small 5-day conference on the foundations of quantum mechanics with a very limited audience. There were only 6 main participants, Yakir Aharonov, P. A. M. Dirac, Wendell Furry, Boris Podolsky, Nathan Rosen, and Eugene Wigner plus a limited audience of 15 other prominent physicists and philosophers.[18] On the first day of the conference in the middle of a discussion of wave function collapse, Podolsky asked Rosen if he would say something about Everett's ideas. There followed a brief discussion during which it was suggested that Everett fly from Washington (to Cincinnati) to discuss his ideas in person. This happened on the following day and a lively discussion ensued. If 5 years after receiving my PhD I had been summoned to appear before such an august panel, it would have been one of the highlights of my career. Evidently, most of these prominent physicists did disregard Everett's interpretation, as Tegmark noted, but it seems clear that his ideas did receive a fair hearing.

It should be clear that from my pragmatic, experimentalist point of view, I am much more confortable with Bohr's Copenhagen interpretation than I am with Everett's universal wave function. For me, Everett's interpretation adds little to my understanding of quantum mechanics and cosmology. This is not to say that one should not attempt to make use of a quantum mechanical model to help us understand the early universe. Whether or not this makes use of a "universal wave function" remains to be seen. However, to simply transplant the wave function and Schrödinger equation from atomic physics and apply it to the entire universe is not well motivated. If one rejects the notion that the wave function has an ontic reality, as do I, then at the very least one should clearly specify the empirical meaning of the universal wave function, i.e., how it is used in conjunction with observations, before employing it in this context. Also, as discussed above, I see no reason why observers should be included in such a wave function. It seems to me that it is neither necessary nor useful to do so. To be clear, I am not suggesting that quantum phenomena are absent in the early universe but rather that one might have to modify our current formalism to include them. For me, it is the empirical concept of the *quantum of action* ($h$) that is the key construct of quantum mechanics, not the wave function or Hilbert state vector (Boughn 2018)[19], and it seems likely that this should be the starting point for any extension of the domain of quantum theory. When Wheeler's support for Everett's interpretation began to wane, he specifically mentioned the importance of the quantum (of action) that he felt Everett had neglected. In 1979 he proclaimed (Byrne 2010, p. 332)

> Imaginative Everett's thesis is, and instructive, we agree. We once
> subscribed to it. In retrospect, however, it looks like the wrong track.
> First, this formulation of quantum mechanics denigrates the quantum.
> It denies from the start that the quantum character of Nature is any clue
> to the plan of physics. Take this Hamiltonian for the world, that
> Hamiltonian, or any other Hamiltonian, this formulation says. I am in
> principle too lordly to care which, or why there should be any Hamiltonian
> at all. You give me whatever world you please, and in return I give you
> back many worlds. Don't look to me for help in understanding this
> universe.

---

[18] This was the same conference from which I took Furry's and Wigner's above quoted remarks.

Anton Zeilinger (1996), who suggested "that the austerity of the Copenhagen interpretation should serve as a guiding principle in a search for deeper understanding [of quantum mechanics]", also points to the importance of the quantum of action in any new formulation:

> If…we accept that there might be a problem of proper philosophical
> foundation in quantum mechanics, the question arises as to what the
> new paradigm should look like, what its features should be. Here, it is
> certainly helpful to investigate which features differentiate the new
> theory from the old one. Of course, the quantum of action is the first to
> catch the eye, especially the fact that there is a universal smallest action
> which can be exchanged in a physical process. I propose that this fact,
> which originates from experiment and which is integrated in theory,
> should actually follow from the new paradigm.

Finally, while it's clear that I don't agree with Everett's interpretation of quantum mechanics, many physicists have maintained that his thesis has provided inspiration in the pursuit of new areas of research in physics. Many of these instances are discussed in Byrne's 2010 book. According to Byrne, Dieter Zeh, Wojciech Zurek, Erich Joos, Murray Gell-Mann, Stephen Hawking, and James Hartle all credit Everett with inspiring

---

[19] In fact, it is possible to derive the Schrödinger equation from Heisenberg's empirical indeterminacy relation, $\delta x \delta p \sim h$. (Hall & Reginatto 2002, Boughn & Reginatto 2018b)

them to think beyond the limitations on understanding imposed by the Copenhagen interpretation and the collapse postulate. Zeh, Zurek, and Joos were the prime movers behind decoherence theory, an important step in solving the preferred basis problem. David Deutsch, a pioneer in the field of quantum information and quantum computing, met Everett when he was a graduate student of Wheeler's at the University of Texas, a meeting from which he found inspiration. At the 1989 Santa Fe conference, *Complexity, Entropy, and the Physics of Information*, "Everett was widely credited as bringing information theory to bear on quantum mechanics." (Byrne 2010, pp. 370-2) At the same meeting Hartle and Gell-Mann credited Everett with suggesting how to apply quantum mechanics to cosmology. "They considered their 'decohering sets of histories' theory as an 'extension' of his work." Another conference participant, Jonathan Helliwell, "…explained that cosmologists owe Everett a debt for opening the door to a completely quantum universe." So whether one considers Everett's interpretation as being an insightful extension of quantum mechanics or a misguided attempt to solve non-existent problems (wave function collapse and the classicality of measurements), as I do, there can be no doubt that Everett's theory has inspired others who have made important contributions to physics (and philosophy).